\documentclass[authoryear]{elsarticle}
\usepackage[a4paper, total={6in, 10in}]{geometry}
\usepackage{booktabs}
\usepackage{float}
\usepackage{amsmath}
\usepackage{amsthm}
\usepackage{ams
math,amssymb,amsfonts,mathrsfs,hyperref,microtype, amsthm}
\usepackage{graphicx}
\usepackage{graphicx,mathabx}
\usepackage{enumitem}  
\graphicspath{ {images/} }
\numberwithin{equation}{section}

\newtheorem{example}{Example}
\newtheorem*{example*}{Example}

\newtheorem{theorem}{Theorem}[section]

\newcommand{\BS}{\boldsymbol}

\newcommand{\Rmnum}[1]{\expandafter\@slowromancap\romannumeral #1@}

\journal{}
\makeatletter
\def\ps@pprintTitle{%
   \let\@oddhead\@empty
   \let\@evenhead\@empty
   \def\@oddfoot{\reset@font\hfil\thepage\hfil}
   \let\@evenfoot\@oddfoot
}

\begin{document}

\begin{frontmatter}
\author[a]{Helmi Shat\corref{cor1}}
\ead{hshat@ovgu.de}
\address[a]{\small Institute for Mathematical Stochastics, Otto-von-Guericke University Magdeburg, \\ \small PF 4120,
39016 Magdeburg, Germany }

 \title{Optimal Design of Stress Levels in Accelerated Degradation Testing for 
Multivariate Linear Degradation Models}

\begin{abstract}
In recent years, more attention has been paid prominently to accelerated degradation testing in order to characterize accurate estimation of reliability properties for systems that are designed to work properly for years of even decades. 
In this paper we propose optimal experimental designs for repeated measures accelerated degradation tests with competing failure modes that correspond to multiple response components. The observation time points are assumed to be fixed and known in advance. The marginal degradation paths are expressed using linear mixed effects models. The optimal design is obtained by minimizing the asymptotic variance of the estimator of some quantile of the failure time distribution at the normal use conditions. Numerical examples are introduced to ensure the robustness of the proposed optimal designs and compare their efficiency with standard experimental designs.
\end{abstract}

\begin{keyword}
Accelerated degradation test\sep multivariate linear mixed effects model\sep the multiplicative algorithm \sep locally $c$-optimal design.

\end{keyword}

\end{frontmatter}

\section {Introduction}
Due to the evolutionary improvements of current industrial technologies, suppliers nowadays are obliged to manufacture highly reliable products in order to compete in the industrial market. Consequently, the reliability estimation of theses products using ALTs might be inefficient as these products are designed to operate without failure for years or even tens of years. Hence, accelerated degradation tests (ADT) is suggested in order to give estimations in relatively short periods of time of the life time and reliability of these systems. For example, \citep{meeker1998accelerated} explained the
 connection between accelerated degradation reliability models and failure-time reliability models. The authors use approximate maximum
 likelihood estimation to estimate model parameters from the underlying mixed-effects
 nonlinear regression model where  simulation-based methods are utilized to compute confidence intervals for
 a certain quantile of the failure time distribution. 
The degradation process in complicated systems may occur due to multiple operating components, where theses components may be independent or have a certain level of interaction.
Hence, ADT in the presence of competing failure modes is an important reliability area to be addressed. Hence, the study of the statistical inference of ADT with competing failures is of great significance and have been considered by many authors, see \citep{duan2018bivariate}. For instance, \citep{haghighi2014accelerated} presented a step-stress test in the presence of competing risks and using degradation
measurements where the underlying degradation process is represented with a
concave degradation model under the assumption that the intensity functions corresponding to competing risks depend only on the level of degradation. In order to obtain the maximum likelihood 
estimates of intensity functions at normal use conditions, the author extrapolates the information from step-stress test at
high level of stress through a tempered failure rate model. For linear models with nuisance parameters, \citep{filipiak2009optimal} gave relationships
between Kiefer optimality of designs in univariate models and in their multivariate extensions
with known and partially known dispersion matrices. With an application in plastic
substrate active matrix light-emitting diodes, \citep{7160792} proposed a modeling approach to
simultaneously analyze linear degradation data and traumatic
failures with competing risks in step stress ADT experiment. In their research, the authors investigate
the convergence criteria with a power law failure rate under
step-stress ADTs. Additionally, they confirm
the asymptotic properties of the maximum likelihood estimates of the parameters for the
proposed model. 
With an application to an electro-mechanical system, \citep{son2011reliability} derived system performance reliability prediction methods considering multiple competing failure modes. The author assumes that the degradation process occurs in terms of a dependence relation between system functionality and
system performance. 
\citep{wang2015optimal} utilized Monte Carlo simulation to derive a cost-constrained optimal design for a constant
stress accelerated degradation test with multiple stresses and multiple degradation measures. The authors assume that the  degradation measures follow
multivariate normal distribution with an application in a pilot-operated safety valve. In accordance with the work of \citep{wang2015optimal}, \citep{wang2017optimal} obtained also an optimal design of step-stress accelerated
degradation test with multiple s tresses
and multiple degradation measures with an application in rubber sealed O-rings. \citep{shat2019experimental} introduced $c$-optimal design for accelerated gedradation testing under competing response components where the marginal responses correpond to linear mixed-effects models along with Gamma process models.
 Furthermore, \citep{8091007} proposed $D$-, $A$- and $V$-optimal designs of ADT with competing failure modes for products suffering
from both degradation failures and random shock failures. 
The theory of optimal designs of experiments for multivariate models is well developed in the mathematical context of approximate design, see \citep{krafft1992d}. For instance, \citep{mukhopadhyay2008comparison} discussed response surface designs for multivariate generalized linear models (GLMs) considering a special case of the
bivariate binary distribution. In order to assess the quality
of prediction associated with a given design, the authors utilize the mean-squared error of prediction matrix. \citep{markiewicz2007optimal} discussed the optimality of an experimental design under the multivariate linear models with a known or unknown dispersion matrix. The authors utilize Kiefer optimality to derive optimal designs for these linear models. 
 \citep{dror2006robust} proposed a simple heuristic for constructing robust experimental designs for multivariate generalized
linear models. The authors incorporate a method of clustering a set of local experimental designs to derive local $D$-optimal designs. \citep{gueorguieva2006optimal} addressed the problem of designing pharmacokinetic experiments in multivariate response
situations. The authors investigate a number of optimisation algorithms, namely
simplex, exchange, adaptive random search, simulated annealing and a hybrid, to obtain locally $D$-optimal designs. \citep{schwabe1996optimum} treated in his monograph the theory of optimal designs for multi-factor models and provides an excellent review on the optimal design theory, i.e. the optimality criteria and the general equivalence theorems, up to that time. In addition, the author gives a comprehensive overview of characterisations of optimal designs for various classes of multi-factor designs in terms of the underlying interaction structure, i.e. for complete product-type interactions, no interactions and partial interactions. Considering random intercept models, \citep{schmelter2008optimal} derived $D$-optimal designs for single and multiple treatments situations. The authors show that in a multi-sample situation the variability of the intercept has substantial influence on the choice of the optimal
design.

The rest of this article is organized as follows. In Section \ref{multivariaterandom234589}, we formulate a multivariate degradation path on the basis of marginal linear mixed effects models (LMEMs). Section \ref{infromationnnnnn345566} is devoted to characterize the possibly estimated parameter vector, the resulting information matrix, and the proposed approximate design for the optimization. The considered optimality criterion for deriving $c$-optimal design based on the failure time distribution is introduced in Section \ref{optimalityxcvbn678}. Section \ref{Numerical Examples} addresses two numerical example under two different testing conditions where the robustness of the proposed optimal designs along with their efficiencies were investigated in comparison to some standard experimental design. Finally, we summarize with some concluding remarks in Section \ref{conclusion}.  All numerical computations were made by using the R programming language\citep{R}.

\section {Model description}
\label{multivariaterandom234589}
In this section, we introduce a formulation of a degradation model with $r$ response components where we assume for each component a LMEM similar to the model presented in \citep{shat2021experimental}. Further, in accordance with \citep{shat2019experimental}, the $r$ components are assumed to be independent within testing units.
Each of these components is observed under a value $\mathbf x$ of the experimental stress variable(s). The stress variable(s) is defined over the design region $\mathcal X$ and kept fixed for each unit throughout the degradation process, but may differ from unit to unit. 
 The number $k$ of measurements and the time points are the same for all individuals $i = 1, ..., n$.  
The measurements $y_{i jl}$, which are realizations of random variables $Y_{i jl}$ at response component $l$, are described by a hierarchical model. For each unit $i$ the observation $Y_{i jl}$ for response component $l$ at time point $t_j$ is given by
\begin{equation} 
\label{multivariate-modelindividualresponse}
Y_{i jl} = \mu_{il}(\mathbf{x}_i, t_j) + \varepsilon_{i jl} ,
\end{equation}
where $\mu_{il}(\mathbf{x}, t)$ is the mean degradation of unit $i$ at response component $l$ and time $t$, when stress $\mathbf{x}$ is applied to unit $i$, and $\varepsilon_{i jl}$ is the associated measurement error at time point $t_j$. The measurement error $\varepsilon_{i jl}$ is assumed to be independent from $\mathbf{x}$ and $t$, and normally distributed with zero mean and error variance $\sigma_{\varepsilon,l}^2>0$ ($\varepsilon_{i jl} \sim \mathrm{N}(0,\sigma_{\varepsilon,l}^2)$). The mean degradation $\mu_{il}(\mathbf{x}, t)$ is assumed to be given by a linear model equation in the stress variable $\mathbf{x}$ and time $t$,
\begin{equation} 
	\label{multivariate-mean-degradation-unit}
	\mu_{il}(\mathbf{x}_i, t_j) = \sum_{q= 1}^{p_l} \beta_{il q} f_{lq}(\mathbf{x}_{i}, t_j) 
	= \mathbf{f}_l(\mathbf{x}_{i}, t_j)^{T} \BS{\beta}_{il}  
\end{equation}
where $\mathbf f{_l}( { \mathbf x},t)= (f_{l1}(\mathbf{x}, t), ..., f_{lp_l}(\mathbf{x}, t))^T$ is the $p_{l}$-dimensional vector of regression functions $f_{lq}(\mathbf{x}, t)$ in both the stress variable(s) $\mathbf{x}$ and the time $t$ considering the $l$th response component and denote by $\boldsymbol{\beta}_{il}=({\beta}_{il1},...,{\beta}_{ilp{_l}})^T$ the $p_l$-dimensional vector of unit specific parameters $\beta_{il s}, s=1,...,p_l,$ at response component $l$. 
Denote by $\textbf g_{{l}}(t  )$ the $q_{l}$-dimensional random effect regression function which only depends on the time $t$, 
and by $\BS{\gamma}_{il} = (\gamma_{il 1}, ..., \gamma_{il q_{l}})^T$ the $q_{l}$-dimensional vector of unit specific deviations $\gamma_{i ls} = \beta_{il s} - \beta_{l s}$, $s = 1, ..., q_{l}$, from the corresponding aggregate parameters. Hence, $\boldsymbol{\gamma}_{il}$ has $q_{l}$-dimensional multivariate normal distribution with zero mean and variance-covariance matrix $\BS{\Sigma}_{\gamma_ l}$ ($\BS{\gamma}_{il} \sim \mathrm{N}(\mathbf{0}, \BS{\Sigma}_{\gamma_ l})$) where $\BS\Sigma_{\gamma_{_{l}}}$ is the corresponding $q_{l}\times q_{{l}}$ positive definite variance covariance matrix.
Assuming that  $\textbf g_{{l}}$ is in the span of $\mathbf f{_l}$,
i.e. $\textbf g_{{l}} = \textbf Q_l \mathbf f{_l}$ for some $q_l \times p_l$ matrix $\textbf Q_l$
such that $\textbf Q_l^T \BS\gamma_{il} = \boldsymbol\beta_{il} - \boldsymbol\beta_l$, the model \eqref{multivariate-modelindividualresponse} can be rewritten for unit $i$ as
\begin{equation}
\label{etrkewjht43876ztrjgnvösfd} 
 Y_{i jl} =  \mathbf{f}_{l}(\mathbf{x}_{i}, t_j)^{T} \BS{\beta}_{l}+\mathbf g_{{l}}(t_j  )^T{\BS\gamma}_{il} + \varepsilon_{i jl}.
\end{equation}
Let $\BS t=(t_1,...,t_k)^T$ be the $k$-dimensional time points of measurements within units which is fixed in advance and is not under disposition of the experimenter. Further, 
denote by $\textbf F_{_l}({\mathbf x_{i}},\BS t)=\big(\mathbf f_{_l}({ \mathbf x_{i}},t_1), ... , \mathbf f_{_l}({ \mathbf x_{i}},t_k)\big)^{T}$ the $k\times p_l$ fixed effect design matrix for the marginal response component $l$ of unit $i$. In vector notation the $k$-dimensional vector $\mathbf{Y}_{il} = (Y_{i 1l},...,Y_{i kl})^T$ can be represented as 
\begin{equation} 
\label{vector-notation-y-il-random-effect}
\textbf{Y}_{il}= \textbf F_{_l}({\mathbf x_{i}},\BS t)\boldsymbol{\beta}_{l} +\mathbf G_{{l}}(\BS t  ){\BS\gamma}_{il}+\boldsymbol \varepsilon_{il}
\end{equation}
where $\mathbf G_{{l}}(\BS t  ) =\left(\mathbf g_{{l}}(t_1  ), ... , \mathbf g_{{l}}(t_k  )\right)^T$ is the $k\times q_{l}$ random effects design matrix. 
The $k$-dimensional vector $\boldsymbol \varepsilon_{il}$ is normally distributed as $\boldsymbol \varepsilon_{il}\sim \mathrm{N}(\textbf 0_k, \sigma_\varepsilon^2\mathbf I_k)$, and $\mathbf I_k$ refers to the $k$-dimensional identity matrix. Hence, the $k$-dimensional vector of observations  $\textbf Y_{il}$ has a multivariate normal distributions as $\textbf Y_{il} \thicksim \mathrm{N}(\textbf F_{_l}({\mathbf x_{i}},\BS t)\boldsymbol{\beta}_{l}, \textbf V_l)$
where 
$\textbf V_l = \mathbf G_{{l}}(\BS t  )\BS\Sigma_{\gamma_l} \mathbf G_{{l}}(\BS t  )^T+ \sigma_\varepsilon^2 \textbf I_k $.
Further, the random effects $\BS\gamma_{il}$ as well as the measurement errors $\boldsymbol \varepsilon_{il}$ of the components $\mathbf Y_{il}$  are assumed to be independent within units, which implies independence of the components $\mathbf Y_{il}$ themselves within units. Hence, the per unit random effects parameter vector $\boldsymbol{\gamma}_{i} =(\boldsymbol{\gamma}_{i1}^{T}, ... , \boldsymbol{\gamma}_{ir}^{T})^{T}$ is normally distributed with zero mean 
and a covariance matrix $\BS\Sigma_\gamma=\textrm{diag}\big(\BS\Sigma_{\gamma_{ l}}\big)_{l=1, ...,r}$ where $q=\sum_{l=1}^r q_{l}$.
 Denote $\boldsymbol \varepsilon_{i} = (\boldsymbol \varepsilon_{{i1}}^{T} , ... , \boldsymbol \varepsilon_{{ir}}^{T})^{T}$ as the cumulative vector of random errors
which is considered to be normally distributed with mean zero and variance covariance matrix $\sigma_\varepsilon^2\mathbf I_{kr}$. 
Hence, the stacked $kr$-dimensional response vector $\textbf{Y}_{i}=(\textbf{Y}_{i1}^T, ... , \textbf{Y}_{ir}^T)^T$ is given by
\begin{equation} 
\label{stacked-kr-response-randomeffect}
\textbf{Y}_{i} = \textbf F({\mathbf x_{i}},\BS t)\boldsymbol{\beta}+ \mathbf G\BS{\gamma}_{i} + \boldsymbol \varepsilon_{i}
\end{equation}
where $ \mathbf G=\mbox{diag}(\mathbf G_{{l}}(\BS t  ) )_{l=1, ...,r}$ is the $ kr\times  q$ block diagonal random effects design matrix, $\textbf F({\mathbf x_{i}},\BS t)=\mbox{diag}\big(\textbf F_{_l}({\mathbf x_{_{i}}},\BS t)\big)_{l=1, ...,r}$ is the $kr\times p$ fixed effect design matrix, and $\boldsymbol{\beta} =(\boldsymbol{\beta}_{1}^{T}, ... , \boldsymbol{\beta}_{r}^{T})^{T}$ refers to the $p$-dimensional overall vector of fixed effects parameters for $r$ response components where $p=\sum_{l=1}^r p_{_l}$.
Then $\mathbf{Y}_i$ is $kr$-dimensional multivariate normally distributed with mean $\textbf F({\mathbf x_{i}},\BS t)\boldsymbol{\beta}$ and variance covariance matrix $\mathbf{V}=\mathbf G{\BS{\Sigma}}_\gamma\mathbf G^T + \sigma_\varepsilon^2\mathbf I_{kr}$. Further, $\mathbf G{\BS{\Sigma}}_\gamma\mathbf G^T=\mbox{diag}(\mathbf G_l\BS\Sigma_{\gamma_{ l}}\mathbf G_l^T)$ and, hence, $\mathbf{V} = \mbox{diag}(\mathbf{V}_l)$, which illustrates the independence of $\mathbf Y_{il}$ within units.
It can be noted that  the variance covariance matrix $\mathbf{V}$ is not affected by the choice of the stress level $\mathbf{x}_i$ and, hence, equal for all units $i$.
For the observations of all $n$ independent units the stacked $n k r$-dimensional response vector $\mathbf{Y}=(\mathbf{Y}_{1}^T, ... , \mathbf{Y}_{n}^T)^T$ can be expressed as
\begin{equation}
   	\label{multivariate-full-model-random-effect} 
	\mathbf{Y} = \mathbf{F}\BS\beta + (\mathbf{I}_n \otimes {\mathbf{G}})\,{\BS\gamma} + \BS\varepsilon ,
\end{equation} 
where $ \mathbf{F} =( \textbf F({\mathbf x_{1}},\BS t)^T,...,\textbf F({\mathbf x_{n}},\BS t)^T)^T$ is the $nkr \times p$ design matrix for the stress variables across units, ${\BS\gamma} = ({\BS\gamma}_1^T, ..., {\BS\gamma}_n^T )^T$ is the $n q$-dimensional stacked parameter vector of random effects. The vector ${\BS\varepsilon} = ({\BS\varepsilon}_1^T, ..., {\BS\varepsilon}_n^T )^T$ is the $n kr$-dimensional stacked vector of random errors which is normally distributed with mean zero and variance covariance matrix $\sigma_\varepsilon^2\mathbf{I}_{nkr}$ ($\BS \varepsilon \sim \mathrm{N}(\mathbf{0}, \sigma_\varepsilon^2\mathbf{I}_{nkr}$)
and the vector ${\BS\gamma}\sim \mathrm{N}(\mathbf{0},\mathbf{I}_n\otimes {\BS{\Sigma}}_\gamma)$ of all random effects is multivariate normal. In total, the vector $\mathbf{Y}$ of all observations is $n kr$-dimensional multivariate normal, $\mathbf{Y}\sim \mathrm{N}(\mathbf{F}\BS\beta, \mathbf{I}_n \otimes \mathbf{V})$. 
 \section{Estimation, information and design}
\label{infromationnnnnn345566}
Under the distributional assumptions of normality for both the random effects and the measurement errors the model parameters may be estimated by means of the maximum likelihood method.
Denote by $\BS\theta = (\BS{\beta}^T, \BS{\varsigma}^T)^T$ the vector of all model parameters where $\BS\varsigma$ indicates the variance covariance parameter vector related to $\sigma_\varepsilon^2$ and $\BS\Sigma_\gamma$. 
The log-likelihood for the current model is given by
\begin{equation}
	\label{logggggglikelihooood}
	\ell(\BS{\theta}; \mathbf{y}) = 
	- {\textstyle{\frac{n kr}{2}}} \log(2\pi) 
	- {\textstyle{\frac{n}{2}}} \log(\det(\mathbf{V})) 
	- {\textstyle{\frac{1}{2}}} (\mathbf y - \mathbf{F} \BS{\beta})^T (\mathbf{I}_n \otimes \mathbf{V})^{-1} (\mathbf{y} - \mathbf{F} \BS{\beta}) ,
\end{equation}
where the variance covariance matrix $\mathbf{V} = \mathbf{V}(\BS\varsigma)$ of measurements per unit depends only on $\BS\varsigma$.
The maximum likelihood estimator of $\BS\beta$ can be calculated as
\begin{eqnarray} 
	\widehat{\BS{\beta}} 
	&=&
	(\mathbf{F}^T (\mathbf{I}_n \otimes \widehat{\mathbf{V}})^{-1} \mathbf{F})^{-1} \mathbf{F}^T (\mathbf{I}_n \otimes \widehat{\mathbf{V}})^{-1} \mathbf{Y}
	\label{multivaiate-eq-maxlik-beta}
\end{eqnarray}
if $\mathbf{F}$ is of full column rank $p$, and $\widehat{\mathbf{V}}=\mathbf{V}(\widehat{\BS\varsigma})$, where $\widehat{\BS\varsigma}$ is the maximum likelihood estimator of $\BS\varsigma$. We note further that $\widehat{\BS{\beta}}$ can be represented by
\begin{eqnarray} 
	\widehat{\BS{\beta}} 
	&=&
\left (\sum_{i=1}^{n} \mathbf F(\mathbf x_i,\mathbf t)^T \widehat{\mathbf V}^{-1} \mathbf F(\mathbf x_i,t)\right )^{-1} \sum_{i=1}^{n} \mathbf F(\mathbf x_i,\mathbf t)^T \widehat{\mathbf V}^{-1} \mathbf Y_i
	\label{multivaiate-eq-maxlik-beta}
\end{eqnarray}


In general, the Fisher information matrix is defined as the variance covariance matrix of the score function $\mathbf{U}$ which itself is defined as the vector of first derivatives of the log likelihood with respect to the components of the parameter vector $\BS{\theta}$.
In particular, let $\mathbf{U} = (\frac{\partial}{\partial \theta_1} \ell(\BS\theta; \mathbf{y}), ..., \frac{\partial}{\partial \theta_q} \ell(\BS\theta; \mathbf{y}))^T$, where $q$ is the dimension of $\BS\theta$.
Then for the full parameter vector $\BS\theta$ the Fisher information matrix is defined as $\mathbf{M}_{\BS\theta} = \mathrm{Cov}(\mathbf{U})$, where the expectation is taken with respect to the distribution of $\mathbf{Y}$.
The Fisher information can also be computed as minus the expectations of the second derivatives of the score function $\mathbf U$, i.\,e.\ $\mathbf{M}_{\BS\theta} = - \mathrm{E}\left(\frac{\partial^2}{\partial \BS\theta \partial \BS\theta^T} \ell(\BS\theta; \mathbf{y})\right)$. 
Under common regularity conditions the maximum likelihood estimator $\widehat{\BS\theta}$ of $\BS\theta$ is consistent and asymptotically normal with asymptotic variance covariance matrix equal to the inverse $\mathbf{M}_{\BS\theta}^{-1}$ of the Fisher information matrix $\mathbf{M}_{\BS\theta}$.
To specify the Fisher information matrix further, denote by $\mathbf{{M}}_{\BS\beta} = - \mathrm{E}\left(\frac{\partial^2}{\partial \BS\beta \partial \BS\beta^T} \ell(\BS\theta; \mathbf{y})\right)$, $\mathbf{M}_{\BS\varsigma} = - \mathrm{E}\left(\frac{\partial^2}{\partial \BS\varsigma \partial \BS\varsigma^T} \ell(\BS\theta; \mathbf{y})\right)$, $\mathbf{M}_{\BS\beta \BS\varsigma} = - \mathrm{E}\left(\frac{\partial^2}{\partial \BS\beta \partial \BS\varsigma^T} \ell(\BS\theta; \mathbf{y})\right)$ and $\mathbf{{M}}_{\BS\varsigma \BS\beta} = \mathbf{{M}}_{\BS\beta \BS\varsigma}^T$ the blocks of the Fisher information matrix corresponding to the second derivatives with respect to $\BS\beta$ and $\BS\varsigma$ and the mixed derivatives, respectively. 
The mixed blocks can be seen to be zero due to the independence property that arises for the normal distribution, and the Fisher information matrix is block diagonal,
\begin{equation}
	\label{multivariate-info-block}
	\mathbf{M}_{\BS{\theta}}
	= \left( 
		\begin{array}{cc}
			\mathbf{M}_{\BS\beta} & \mathbf{0}
			\\
			\mathbf{0} & \mathbf{M}_{\BS\varsigma}
		\end{array}
	\right) .
\end{equation}  
Moreover, the block $\mathbf{M}_{\BS\beta}$ associated with the aggregate location parameters $\BS\beta$ turns out to be the inverse of the variance covariance matrix for the estimator $\widehat{\BS\beta}$ of $\BS\beta$, when $\mathbf{V}$ is known.
Actually, because the Fisher information matrix for $\BS\theta$ is block diagonal, the inverse $	\mathbf{M}_{\BS\beta}^{-1}$ of the block associated with $\BS\beta$ is the corresponding block of the inverse of $\mathbf{M}_{\BS\theta}$ and is, hence, the asymptotic variance covariance matrix of $\widehat{\BS{\beta}}$.
Accordingly the asymptotic variance covariance matrix for estimating the variance parameters $\BS\varsigma$ is the inverse of the block $\mathbf{M}_{\BS\varsigma}$. In the following we will refer to $\mathbf{M}_{\BS\beta}$ and $\mathbf{M}_{\BS\varsigma}$ as the information matrices for $\BS\beta$ and $\BS\varsigma$, respectively, for short.
The particular form of $\mathbf{M}_{\BS\varsigma}$ will be not of interest here.
It is important to note that $\mathbf{M}_{\BS\varsigma}$ does not depend on the settings $\mathbf{x}_1, ..., \mathbf{x}_n$ of the stress variable in contrast to the information matrix $\mathbf{M}_{\BS\beta}$ of the aggregate location parameters $\BS\beta$.

The quality of the estimates will be assessed in regards to the information matrix and, hence, depends on the settings of the stress variable as well as the time points of measurements. When these variables are controlled by the experimenter, then their choice will be called the design of the experiment.
As mentioned earlier the time points of measurements within units is fixed in advance and is not considered for the optimization process.
Then only the settings $\mathbf{x}_1, ..., \mathbf{x}_n$ of the stress variable $\mathbf{x}$ can be adjusted to the units $i = 1, ..., n$.
Their choice $(\mathbf{x}_1, ..., \mathbf{x}_n)$ is then called an ``exact'' design, and their influence on the performance of the experiment is indicated by adding them as an argument to the information matrices, $\mathbf{M}_{\BS\theta}(\mathbf{x}_1, ..., \mathbf{x}_n)$ and $\mathbf{M}_{\BS\beta}(\mathbf{x}_1, ..., \mathbf{x}_n)$, where appropriate.
It should be noted that $\mathbf{M}_{\BS\varsigma}$ does not depend on the design for the stress variable.
For $\mathbf x_1,...,\mathbf x_n$ a general form of the information matrix with respect to the location parameter vector $\BS{\beta}$ is defined as
\begin{equation}
\begin{split}
\textbf{{M}}_{\BS{\beta}}(\mathbf x_1,...,\mathbf x_n)&=\sum_{i= 1}^n \textbf F({\mathbf x_{i}},\BS t)^{T}  \mathbf{V}^{-1}
\textbf F({\mathbf x_{i}},\BS t).
\end{split}
\end{equation}
By the independence of the components the information matrix $\textbf{{M}}_{\BS{\beta}}$ decomposes into its marginal counterparts
$\textbf{{M}}_{\BS{\beta}} = \mbox{diag}(\textbf{{M}}_{\BS{\beta}_l})$ where $\textbf{{M}}_{\BS{\beta}_l} = \sum_{i=1}^{n} \textbf F_l({\mathbf x_{i}},\BS t)^{T}  \mathbf{V}_l^{-1}
\textbf F_l({\mathbf x_{i}},\BS t)$.
In general, 
${\boldsymbol{\beta}}$ can be estimated by MLE or, more usually, by restricted maximum likelihood (REML) \citep{debusho2008v} providing that REML has the same asymptotic property as MLE. In addition, the variance covariance matrix of the estimator $\widehat{{\boldsymbol{\beta}}}$ of the location parameters ${{\boldsymbol{\beta}}}$ can be asymptotically approximated by the inverse of the information matrix ${\textbf{{M}}}_{{\BS\beta}}(\mathbf x_1,...,\mathbf x_n).$ 
It can easily be seen that the information matrix ${\textbf{{M}}}_{{\BS\beta}}(\mathbf x_1,...,\mathbf x_n)$ does not depend on the order of the settings but only on their mutually distinct values, $\mathbf{x}_1, ..., \mathbf{x}_m$ say, and their corresponding frequencies $n_1, ..., n_m$, such that $\sum_{i = 1}^m n_i = n$, i.\,e.\ $\mathbf{M}_{\BS\beta} = \sum_{i = 1}^m n_i 
\textbf F({\mathbf x_{i}},\BS t)^{T}  \mathbf{V}^{-1}
\textbf F({\mathbf x_{i}},\BS t)
$.
Finding optimal exact designs is, in general, a difficult task of discrete optimization.
To circumvent this problem we follow the approach of approximate designs propagated by \citep{kiefer1959optimum} in which the requirement of integer numbers $n_i$ of testing units at a stress level $\mathbf{x}_i$ is relaxed.
Then continuous methods of convex optimization can be employed (see e.\,g.\ \citep{silvey1980optimal}) and efficient exact designs can be derived by rounding the optimal numbers to nearest integers.
This approach is, in particular, of use when the number $n$ of units is sufficiently large.
Moreover, the frequencies $n_i$ will be replaced by proportions $w_i=n_i/n$, because the total number $n$ of units does not play a role in the optimization. 
Thus an approximate design $\xi$ is defined by a finite number of settings $\mathbf{x}_i$, $i=1,...,m$, from an experimental region $\mathcal{X}$ with corresponding weights $w_i\geq 0$ satisfying $\sum_{i = 1}^m w_i = 1$ and is denoted by
\begin{equation}
	\label{multivariateeq-design-x}
	\xi = \left(
		\begin{array}{ccc}
			\mathbf{x}_1& ... &\mathbf{x}_{m} 
			\\ 
			w_{1}& ... &w_{m}
		\end{array} 
	\right),
\end{equation}
The corresponding standardized, per unit information matrix is accordingly defined as
\begin{equation}
	\label{multivariate-eq-info_1-xi}
	\mathbf{M}_{\BS\beta}(\xi)
	= \sum_{i = 1}^m w_i\textbf F({\mathbf x_{i}},\BS t)^{T}  \mathbf{V}^{-1}
\textbf F({\mathbf x_{i}},\BS t)
\end{equation}
for the aggregate parameters $\BS\beta$. By the independence of the components $\mathbf{M}_{\BS\beta}(\xi)$ decomposes accordingly
$\mathbf{M}_{\BS\beta}(\xi) = \mbox{diag}(\mathbf{M}_{\BS{\beta}_l}(\xi))$  where $\mathbf{M}_{\BS{\beta}_l}(\xi) =\sum_{i = 1}^m w_i\textbf F_l({\mathbf x_{i}},\BS t)^{T}  \mathbf{V}_l^{-1}
\textbf F_l({\mathbf x_{i}},\BS t)$. For the full parameter vector $\BS\theta$ the standardized, per unit information matrix $\mathbf{M}_{\BS\theta}(\xi)$ is expressed as
\begin{equation}
	\label{multivariate-eq-info_theta-xi}
	\mathbf{M}_{\BS\theta}(\xi)
	= \left(
		\begin{array}{cc}
			 \mathbf{M}_{{\BS\beta}}(\xi)  & \mathbf{0}\\
			 \mathbf{0} & \widetilde{\mathbf{M}}_{\BS\varsigma}
		\end{array}
	\right)
\end{equation}
where now $\widetilde{\mathbf{M}}_{\BS\varsigma} = \frac{1}{n} \mathbf{M}_{\BS\varsigma}$ is the standardized, per unit information for the variance parameters $\BS\varsigma$.
If all $n w_i$ are integer, then these standardized versions coincide with the information matrices of the corresponding exact design up to the normalizing factor $1/n$ and are , hence, an adequate generalization.
In order to optimize information matrices, some optimality criterion has to be employed which is a real valued function of the information matrix and reflects the main interest in the experiment.

\section{Optimal design based on failure times}
\label{optimalityxcvbn678}
In accordance with \citep{shat2021experimental} we consider some characteristics of the failure time distribution of soft failure due to degradation. For the analysis of degradation under normal use we further assume that the general model~(\ref{etrkewjht43876ztrjgnvösfd}) is also valid at the normal use condition $\mathbf{x}_u$, i.\,e.
\begin{equation}
	\label{multivariate-eq-degr-path-use-cond}
	\mu_{ul}(t) =\mu_l(\mathbf{x}_{u},t)= \mathbf{f}_l(\mathbf{x}_{u}, t)^{T} \BS{\beta}_{l}+\mathbf g_l(t)^T{\BS\gamma}_{ul} 
\end{equation} 
describes the mean degradation of a future unit $u$ at normal use condition $\mathbf{x}_u$, time $t$ and response component $l$  where $\mu_{ul}$ denotes the degradation path under normal use condition for short. 
Further, denote $\mu_l(t) = \mathrm{E}(\mu_{ul}(t)) = \mathbf{f}_l(\mathbf{x}_{u}, t)^{T} \BS{\beta}_{l}$ as the aggregate degradation path under normal use condition for response component $l$. 
Under the assumption that the $r$ mean degradation paths are strictly increasing over time,
a soft failure at component $l$ is defined as the exceedance of the degradation over a  failure threshold $y_{l0}$ on the basis of the  degradation path.
The marginal failure time $T_l$ under normal use condition is then defined as the first time $t$ the mean degradation path $\mu_{ul}(t)$ reaches or exceeds $y_{l0}$, i.\,e.\ $T_l = \min \{t \geq 0;\, \mu_{ul}(t) \geq y_{l0}\}$.
As the mean degradation path includes the random effect $\gamma_{ul}$, the marginal failure time $T_l$ is random.
In order to express certain characteristics of the failure time distribution, we will describe first the marginal distribution function $F_{T_l}(t) = \mathrm{P}(T_l \leq t)$.
First note that $T_l \leq t$ if and only if $\mu_{ul}(t) \geq y_{l0}$.
Subsequently
\begin{eqnarray}
\label{F-t-marginal}
   F_{T_l}(t) &  = & \mathrm{P}(\mu_{ul}(t) \geq y_{l0})
   \nonumber
   \\
   & = & \mathrm{P}(\mu_l(t) + \mathbf g_l(t)^T{\BS\gamma}_{ul}  \geq y_{l0}) 
   \nonumber
	\\
	& = & \mathrm{P} (-  \mathbf g_l(t)^T{\BS\gamma}_{ul}  \leq \mu_l(t)-  y_{l0}) 
   \nonumber
   \\
	& = & \Phi(h_l(t)) ,
    \label{multivariate-eq-failure-time-distribution}
\end{eqnarray}
where
\begin{equation}
	\label{multivariate-eq-h-tau}
	h_l(t) = \frac{ \mu_l(t) -  y_{l0}}{\sigma_{ul}(t)} ,
\end{equation}
$\sigma_{ul}^2(t) =  \mathbf g_{{l}}(t  )^T \BS\Sigma_{\gamma_l}  \mathbf g_{{l}}(t  )$ is the variance $\mu_{ul}(t)$ at time $t$ and $\Phi$ indicates the standard normal distribution function. In the particular case of straight lines for the mean degradation paths, i.e. $\mathbf g_{l}( t )=(1,t)^T, \, l=1,..,r,$ the variance covariance matrix is given by $\BS{\Sigma}_{\gamma_l} =
\left(\begin{array}{cc}
	\sigma_{l1}^2 & \rho_l \sigma_{l1} \sigma_{l2}
	\\
	\rho_l \sigma_{l1} \sigma_{l2} & \sigma_{l2}^2
\end{array}\right)$, and, hence, the function $h_l(t)$ specifies to
\begin{equation}
	\label{eq-h-tau-linear}
	h_l(t) = \frac{\mu_l(t)  - y_{l0}}{\sqrt{\sigma_{l1}^{2} + 2\rho_l \sigma_{l1}\sigma_{l2} t + \sigma_{l2}^{2} t^2}} ,
\end{equation}

The joint failure time $T$ is defined consequently as a function, say $\psi$, of the marginal failure times, $T = \psi(T_1,...,T_r).$
For instance, a failure for an $s$-out-of-$r$ system occurs if, at least, $s$ of its  $r$ components exceed their corresponding failure thresholds. Hence, for the special case $1$-out-of-$r$ system, the joint failure time $T$ might be defined as $T = \min(T_1,...,T_r)$ so that a failure of the system occurs if, at least, one of its components fail
.
Quantiles $t_\alpha$ of the joint failure time distribution, i.\,e.\ $\mathrm{P}(T \leq t_\alpha) = \alpha$, are considered for further calculations.
For each $\alpha$ the quantile $t_\alpha$ indicates the time up to which under normal use conditions (at least) $\alpha \cdot 100$ percent of the units fail and (at least) $(1 - \alpha) \cdot 100$ percent of the units survive.
The quantiles $t_\alpha$ are increasing in $\alpha$.
Further, the current standard definition of quantiles is in contrast to the ``upper'' quantiles ($t_{1-\alpha}$) used in  \citep{weaver2014methods} where percentages of failures and persistence are reversed.
Of particular interest is the median $t_{0.5}$ up to which under normal use conditions half of the units fails and half of the units persist ($\alpha = 0.5$).
The quantile $t_\alpha = t_\alpha(\BS\theta)$ is a function of both the location parameters vector $\BS\beta$ as well as the variance parameters $\BS\varsigma$.
Hence, the maximum likelihood estimator of the quantile $t_\alpha$ is given by $\widehat{t}_\alpha = t_\alpha(\widehat{\BS\theta})$ in terms of the maximum likelihood estimator $\widehat{\BS\theta}$ of $\BS\theta$.
The task of designing the experiment will now be to provide an as precise estimate of the $\alpha$-quantile as possible. By the delta-method $\widehat{t}_\alpha$ is seen to be asymptotically normal with  asymptotic variance
\begin{equation}
	\label{multivariate-eq-avar-tau-alpha}
	\mathrm{aVar}(\widehat{t}_{\alpha}) = \mathbf{c}^T \mathbf{M}_{\BS{\theta}}^{-1} \mathbf{c} ,
\end{equation}
where $\mathbf{c} = \frac{\partial}{\partial\BS\theta}t_{\alpha}$ is the gradient vector of partial derivatives of $t_{\alpha}$ with respect to the components of the parameter vector ${\BS\theta}$. 
The asymptotic variance depends on the design of the experiment through the information matrix $ \mathbf{M}_{\BS{\theta}}$ and will be chosen as the optimality criterion for the design.
Considering the independence between $\BS\beta$and $\BS\varsigma$ the overall gradient simplifies to $\mathbf{c} =  (\mathbf{c}_{\BS\beta}^T, \mathbf{c}_{\BS\varsigma}^T)^T$, where
\begin{equation*}
	\label{multivariate-eq-gradient-tau-alpha-beta}
	\mathbf{c}_{\BS\beta} = {\textstyle{\frac{\partial}{\partial\BS\beta}}}t_\alpha(\BS\theta)
\end{equation*}
is the gradient of $ t_\alpha$ with respect to $\BS\beta$ and
\begin{equation*}
	\label{multivariate-eq-gradient-tau-alpha-sigma}
	\mathbf{c}_{\BS\varsigma} = {\textstyle{\frac{\partial}{\partial\BS\varsigma}}}t_\alpha(\BS\theta)
\end{equation*}
is the gradient of $ t_\alpha$ with respect to $\BS\varsigma$ where the particular shape of $\mathbf{c}_{\BS\varsigma}$ does not play a role here, in general.
Due to the block diagonal form of the information matrix in equation (\ref{multivariate-info-block}) the asymptotic variance (\ref{multivariate-eq-avar-tau-alpha}) of $\widehat{t}_\alpha$ specifies to
\begin{equation}
	\label{multivariate-eq-avar-tau-alpha-sum}
	\mathrm{aVar}(\widehat{t}_{\alpha}) =  \mathbf{c}_{\BS\beta}^T \mathbf{M}_{\BS{\beta}}^{-1} \mathbf{c}_{\BS\beta} + \mathbf{c}_{\BS\varsigma}^T \mathbf{M}_{\BS{\varsigma}}^{-1} \mathbf{c}_{\BS\varsigma}
\end{equation}
where the second term in the right hand side  is an additive constant and does not depend on $\xi$.\\
Due to the complexity of deriving an explicit formula of $t_{\alpha}$, the following equality is ensured by the implicit function theorem, see \citep{krantz2012implicit}\\
\begin{equation}
\label{gradieggfgfgf}
\frac{\partial
t_\alpha(\BS\theta)}{\partial{\boldsymbol{\theta}}}=\Bigg(\frac{\partial  F_T(t_{\alpha}(\BS\theta))}{\partial t}\Bigg)^{-1}\frac{\partial  F_T(t_{\alpha}(\BS\theta))}{\partial {\boldsymbol{\theta}}}. \quad\quad
\end{equation}\\
given that $\frac{\partial  F_T(t_{\alpha}(\BS\theta))}{\partial t}$  is a scaling constant that is irrelevant to the design, and the equality is quaranteed in terms of the function $F_T(t(\BS\theta))-\alpha=0\vert_{t=t_\alpha}$. The gradient vector $\mathbf{c}_{\BS\beta} $ can be expressed as $\mathbf{c}_{\BS\beta}=\frac{\partial  F_T(t_{\alpha})}{\partial {\boldsymbol{\beta}}}$ such that $\mathbf c_{\BS\beta}$ decomposes into marginal components
$\mathbf {c}_{\BS\beta} = (\mathbf{c}_{1}^T, ..., \mathbf{c}_{r}^T)^T$
where $\mathbf {c}_{l} = \partial \mu_l(t_\alpha) / \partial \BS\beta_l=c_l\mathbf{f}_{l}(\mathbf{x}_{u}, t_\alpha)$ and $c_l$ is a constant.
 Because the components are assumed to be independent within units, the information matrix $\mathbf{M}_{\BS\beta}(\xi)$ is block diagonal with diagonal entries $\mathbf{M}_{\BS\beta_1}(\xi),..., \mathbf{M}_{\BS\beta_r}(\xi)$ as noted in the previous section. 
Accordingly, based on the optimality criterion defined in equation (\ref{multivariate-eq-avar-tau-alpha-sum}), the gradient vector depends only on the parameter vector ${\BS\beta}$ and
the locally $c$-optimal design $\xi^*$ can be defined by\\
\begin{equation} 
\label{multivariate-111dfdsdfdsfdsffdf}
 \xi^*= \mbox{arg} \min_{ \xi}\left(  \sum_{l=1}^r\mathbf{c}_{l}^T \mathbf{M}_{\BS{\beta}_l}^{-1}(\xi) \mathbf{c}_{l}\right).
\end{equation}
\begin{theorem}
	\label{th-multi}
If all components are described by the same model equations and have the same values for the variance-covariance parameters, i.e. $\mathbf \Sigma_{\gamma_1}=...=\mathbf \Sigma_{\gamma_r} $,\,\, $\mathbf g_1 =...=\mathbf g_r =\mathbf g$, $\mathbf f_1 =...=\mathbf f_r =\mathbf f$, and eventually $\sigma_{\varepsilon,1}^2 = ... = \sigma_{\varepsilon,r}^2$ such that $\mathbf f$, as assumed in \citep{shat2021experimental}, has product type structure $\mathbf f(\mathbf x,t) =\mathbf f^{(1)}(\mathbf x) \otimes \mathbf g(t)$, then the terms in the criterion \eqref{multivariate-111dfdsdfdsfdsffdf} factorize, 
$\sum_{l=1}^{r} \tilde{c}_l\mathbf f^{(1)}(\mathbf x_u)^T\mathbf M^{(1)}(\xi)^{-1}\mathbf f^{(1)}(\mathbf x_u)$
where $\mathbf M^{(1)}(\xi) = \sum_{i=1}^{m} w_i\mathbf f^{(1)}(\mathbf x_i)\mathbf f^{(1)}(\mathbf x_i)^T$ is the (fixed effect) information matrix in the first marginal model related to the stress variable and $\tilde{c}_l = c_l^2 \mathbf g(t_\alpha)^T\mathbf M^{(2)}(\mathbf{t}) \mathbf g(t_\alpha)$ where $\mathbf M^{(2)}=\mathbf{G}_l^T\mathbf V_l^{-1}\mathbf{G}_l,\,l=1,...,r,$ is identical vor all $r$ components.
\end{theorem}

 It should be noted that the assumption of product type structure in $\mathbf f$ guarantees the requirement $\mathbf g$ in span $\mathbf f$ when 1 in span $\mathbf f^{(1)}$, e.g. when the first entry $f_1^{(1)}$ in $\mathbf f^{(1)}$ is constnat 1. On the basis of Theorem \ref{th-multi},   if $\xi$ is optimal for extrapolation at $\mathbf x_u$ in the first marginal model, it is also optimal for estimating $t_\alpha$ in the system. Actually, this holds not only for series system but also for $s$-out-of-$r$ systems in which case the constants $c_l$ are more complicated (see below in Subsection \ref {Numerical Examples}).
In order to assess the influence of the variation of the optimal weights we consider the efficiency of the resulting optimal design optimal design $\xi^*$ when the underlying nominal values are varied.
where the asymptotic efficiency of the design $\xi$ for estimating $t_{\alpha}$ is defined by
\begin{equation}
\label{eff-forboth}
	\mathrm{eff}_{\mathrm{aVar}}(\xi,\BS\beta) = \frac{\mathbf{c}_{\BS\varsigma}^T \widetilde{\mathbf{M}}_{\BS\varsigma}^{-1} \mathbf{c}_{\BS\varsigma}+ \sum_{l=1}^r\mathbf{c}_{l}^T \mathbf{M}_{\BS{\beta}_l}^{-1}(\xi^*) \mathbf{c}_{l}}{\mathbf{c}_{\BS\varsigma}^T \widetilde{\mathbf{M}}_{\BS\varsigma}^{-1} \mathbf{c}_{\BS\varsigma}+ \sum_{l=1}^r\mathbf{c}_{l}^T \mathbf{M}_{\BS\beta_l}^{-1}(\xi) \mathbf{c}_{l}}.
\end{equation}
In view of Theorem \ref{th-multi} it would be helpful to mention that in this situation the efficiency of $\xi$ is inherited from the marginal model: $\mathrm{eff}_{\mathrm{aVar}}>= \mathrm{eff}_{\mathbf x_u}$
where $\mathrm{eff}_{\mathbf x_u}$ is the efficiency for extrpolation at $\mathbf x_u$ in the first marginal model.
\section{Numerical Examples}
\label{Numerical Examples}
In this section we present optimal designs for two examples of accelerated degradation testing.
We consider first an example for a series system in accordance with the work of \citep{shat2021experimental} with full interaction between stress and time variables. We propose further another example for an $s$-out-of-$r$ system with $r$ statistically independent response components under the assumption of partial interaction of explanatory variables with the time variable and identical model equations for all $r$ compoentns. 
 For the latter example, in accordance with the work of \citep{koucky2003exact}, we 
denote by $F_D(t)=\mathrm{P}(T_l \leq t\,\, \forall l \in D) = \prod_ {l \in D} F_{T_{l}}(t)  $ the probability of joint failure of the components in the subset $D\subseteq\{1,...,r\}$. Consequently, the joint failure time distribution function for a $s$-out-of-$r$ system is expressed as,\\
\begin{equation}
\label{eqn:12}
F_T(t) = \sum_{l=0}^{s-1} (-1)^{l}  \binom{l + r - s}{l} \sum_{D: |D|=l+1+r-s} \prod_{d\in D} F_{T_d}(t)
\end{equation}
where, for instance, the serial system occurs for $s=1$.
\begin{example}

\label{example-with-intercept}
We derive in this example a locally $c$-optimal design for the  degradation model in section \ref{multivariaterandom234589} under 
the standardized time plan $\boldsymbol{{t}}=(0, 0.5,1)$, i.e. $k=3$, which is identical for all testing units. 
The degradation is influenced by two standardized accelerating stress variables $\mathbf{x}=(x_1,x_2)$ which are defined over the design region $\mathcal X=[0,1]^2$ and act linearly on the response  with a potential interaction effect associated with $x_1 x_2$. As in the univariate situation described in \citep{shat2021experimental}, for each testing unit $i$, the stress variables are set to $\mathbf{x}_i=(x_{i1},x_{i2})$, and for each component $l$ the response $y_{ilj}$ at time $t_j$ is given by
\begin{equation} 
\begin{split}
\label{example2_1}
 y_{i jl} = &\beta_{il 1} + \beta_{l2} x_{i 1} + \beta_{l3} x_{i 2} + \beta_{l4} x_{i 1} x_{i 2} + \beta_{il 5} t_j + \beta_{l6} x_{i 1} t_j + \beta_{l7} x_{i 2} t_j + \beta_{l8} x_{i 1} x_{i 2} t_j + \varepsilon_{i jl},\\ 
=& \mathbf f_l(\mathbf x_i, t_j)^T\BS \beta_{il}+ \varepsilon_{i jl}\\
\end{split}
\end{equation}
where the vector of regression functions $\mathbf f_l(\mathbf x, t)=\mathbf f(\mathbf x, t)=(1,x_{1},x_{2},x_{1}x_{2},t,x_{1}t,x_{2}t,x_{1}x_{2}t)^T$ is the same for all components $l$ and $\BS \beta_{il}=(\beta_{il 1},\beta_{l 2},\beta_{l 3},\beta_{l 4},\beta_{il 5},\beta_{l 6},\beta_{l 7},\beta_{l 8})^T$. Consequently, it should be further noted that here $\mathbf g_1=...=\mathbf g_r=\mathbf g$ where $g(t)=(1,t)^T$. 
As noted in Section \ref{optimalityxcvbn678} the model equation (\ref{multivariate-mean-degradation-unit}) for the mean degradation paths is also assumed to be valid under normal use condition $\mathbf{x}_u=(x_{u1},x_{u2})$. Hence, the aggregate degradation path under normal use conditions 
is given by
\begin{equation}
\label{marginalfixeddddddd}
\mu_l(t) =  \mathbf{f}_l(\mathbf{x}_{u}, t)^{T} \BS{\beta}_{l}= \delta_{l1}+ \delta_{l2}t 
\end{equation}
where $\delta_{l1}=\beta_{l1}+\beta_{l2}x_{u1}+\beta_{l3}x_{u2}+\beta_{l4}x_{u1}x_{u2}$ and $\delta_{l2}=\beta_{l5}+\beta_{l6}x_{u1}+\beta_{l7}x_{u2}+\beta_{l8}x_{u1}x_{u2}$ are the intercept and the slope of the aggregate degradation path $\mu_l(t)$ under normal use conditions, respectively. 
For the particular case of a series system with two response components, i.e. $r=2$, the joint failure time  distribution function can be expressed as  \\
\begin{equation}
\label{multi-failure-fuinctoin-with-alpha}
 F_T(t)
=\mathrm{P}\Big(\min({T_1},{T_2})\leqslant t\Big)
=1-\Bigg(1-\Phi \Big(\frac  {\delta_{11}+\delta_{12}t - y_{10}   }{   \sigma_1(t)  }\Big)\Bigg)
\times \Bigg(1-\Phi \Big(\frac { \delta_{21}+ \delta_{22} t- y_{20}   }{\sigma_2(t)}\Big)\Bigg).
\end{equation}
where $\sigma_{l}^2(t) = Var(\mu_{ul}(t)) = \mathbf g_l(t)^T\BS{\Sigma}_{\gamma_l} \mathbf g_l(t)$ is the variance function of the mean degradation path of component $l$.
For illustration, the distribution function $F_T(t)$ is plotted in Figure~\ref{joint-failure-distrib-bivariatelinear-withintercept} under the nominal values given in Table~\ref{table:table34df3}, the normal use conditions $x_{u1}=-0.40$ and $x_{u2}=-0.20$, and the failure thresholds $y_{10}=5.4$ and $y_{20}=5.8$. The median failure time $t_{0.5} = 5.2$ is indicated in Figure~\ref{joint-failure-distrib-bivariatelinear-withintercept} by a dashed vertical line.
\begin{figure}
	\centering
		\includegraphics[width=0.38\textwidth]{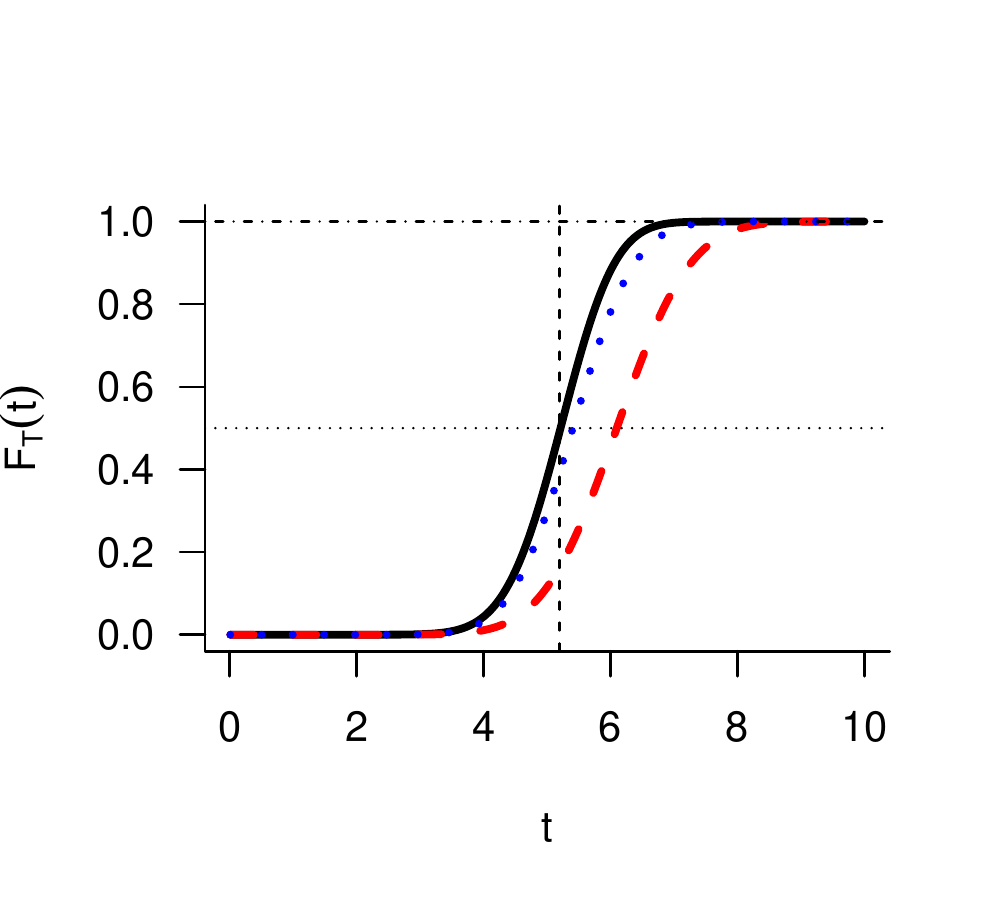}
	\caption{Distribution function $F_T(t)$ (solid line) at the bivariate linear model with random intercept for Example~\ref{example-with-intercept}, dashed line: $F_{T_1}(t)$, dotted line: $F_{T_2}(t)$} 
	\label{joint-failure-distrib-bivariatelinear-withintercept}
\end{figure}
 Consequently, in view of (\ref{gradieggfgfgf}) and (\ref{multi-failure-fuinctoin-with-alpha}) the gradient vector of the parameter vector $\BS\beta$ can be expressed as $\mathbf{c}_{\BS\beta}=({c}_{{1}}\mathbf f(\mathbf x_u,t_{{\alpha}})^T,{c}_{{2}}\mathbf f(\mathbf x_u,t_{{\alpha}})^T)^T$ where the constants ${c}_{{1}}$ and ${c}_{{2}}$ are given by
\begin{eqnarray*}
\label{multivariate-gradient_cb1}
 {c}_{{1}}&=&\phi\left(\frac  {\delta_{11}+\delta_{12}t_{\alpha} - y_{10}   }{\sqrt{\sigma_{11}^2 + 2 \rho_1 \sigma_{11} \sigma_{12} t_{\alpha} + \sigma_{12}^2 t_{\alpha}^2}}\right)\left(1-\Phi\left(\frac { \delta_{21}+ \delta_{22}t_{\alpha} - y_{20}   }{\sqrt{\sigma_{21}^2 + 2 \rho_2 \sigma_{21} \sigma_{22} t_{\alpha} + \sigma_{22}^2 t_{\alpha}^2}}\right)\right),\\
 {c}_{{2}}&=&\phi\left(\frac  {\delta_{21}+\delta_{22} t_{\alpha}- y_{20}   }{\sqrt{\sigma_{21}^2 + 2 \rho_2 \sigma_{21} \sigma_{22} t_{\alpha} + \sigma_{22}^2 t_{\alpha}^2}}\right)\left(1-\Phi\left(\frac { \delta_{11}+\delta_{12}t_{\alpha} - y_{10}   }{\sqrt{\sigma_{11}^2 + 2 \rho_1 \sigma_{11} \sigma_{12} t_{\alpha} + \sigma_{12}^2 t_{\alpha}^2}}\right)\right),
\end{eqnarray*}
where $\phi$ denotes the density of the standard normal distribution.



\begin{table}
	\begin{center}
		\caption{Nominal values of the bivariate linear model in Example \ref{example-with-intercept}.}
		\label{table:table34df3}
		\vspace{5mm}
		\begin{tabular}{c|c|c|c|c|c|c|c}

			 $\beta_{11}$ & $\beta_{12}$ & $\beta_{13}$&$\beta_{14}$ & $\beta_{15}$ & $\beta_{16}$ & $\beta_{17}$& $\beta_{18}$ \\ \hline
$2.30$ & $1.60$ & $1.30$&$0.02$ & $0.70$ & $0.07$ & $0.08$&$0.03$  \\ \hline 

\hline\hline

\hline
$\beta_{21}$ & $\beta_{22}$ & $\beta_{23}$ &$\beta_{24}$ &$\beta_{25}$ & $\beta_{26}$ & $\beta_{27}$& $\beta_{28}$ 
			\\
			\hline
 $2.17$ & $1.10$ &$0.84$ &$0.01$ & $0.80$ 
			& $0.03$& $0.02$ & $0.02$ 
\\
\hline\hline

\hline
$\sigma_{11}^2 = \sigma_{21}^2$ & $\sigma_{12}^2 = \sigma_{22}^2$ &$\rho_1 = \rho_2$& $\sigma^2_\varepsilon$ &

			\\
			\hline
  
$0.36$&$0.10$ &$0.00$& $0.10$\\

		\end{tabular} 
	\end{center}
\end{table}
It can be concluded that the optimal design for each of the two univariate components is also optimal for the joint bivariate model under the condition of same model equations for the response components.
Consequently, the problem has been reduced now to finding an optimal design for any of a univariate model with two explanatory variables. In the model with two interacting stress variables $x_1$ and $x_2$ the marginal model for the combined stress variable $\mathbf{x} = (x_1, x_2)$ is given itself by a product-type structure given both components $x_1$ and $x_2$ are specified as simple linear regressions in their corresponding submarginal models. As depicted in \citep{shat2021experimental} the degradation model in equation\eqref{example2_1} after rearranging terms can be rewritten as a Kronecker product model
\begin{equation}
\label{factorization-bivariate-linear} 
	Y_{i jl} = (\mathbf{f}_{1}(x_{i 1}) \otimes \mathbf{f}_{ 2}(x_{i 2}) \otimes \mathbf{g}(t_j))^{T} \BS{\beta}_l + \mathbf{g}(t_j)^{T} \BS\gamma_{i l}+ \varepsilon_{i jl},
\end{equation}
where $\mathbf{f}_{1}(x_1) = (1, x_1)^T$ and $\mathbf{f}_{2}(x_2) = (1, x_2)^T$ are the marginal regression functions for the stress variables $x_1$ and $x_2$, respectively. 
	Moreover, the experimental region $\mathcal{X} = [0, 1]^2$ for the combined stress variable $\mathbf{x}$ is the Cartesian product of the marginal experimental regions $\mathcal{X}_1 = \mathcal{X}_2 = [0, 1]$ for the components $x_1$ and $x_2$, respectively.
	In this setting the $c$-optimal design $\xi^*$ for extrapolation at $\mathbf{x}_u$ can be obtained as the product $\xi^* = \xi_1^* \otimes \xi_2^*$ of the $c$-optimal designs $\xi_l^*,\,l
=1,2,$  for extrapolation at $x_{u l}$ in the submarginal models (see Theorem~4.4 in \citep{schwabe1996optimum}). 
	
	As specified in \citep{shat2021experimental} the submarginal $c$-optimal designs $\xi_l^*$ assign weight $\tilde w_l^* = |x_{u l}| / (1 + 2|x_{u l}|)$ to $x_l= 1$ and weight $1 - \tilde w_l^* = (1 + |x_{u l}|) / (1 + 2|x_{u l}|)$ to $x_l = 0$.
	Hence, the $c$-optimal design $\xi^* = \xi_1^* \otimes \xi_2^*$ for extrapolation at $\mathbf{x}_u$ is given by
	\[
	\xi^* = \left(
		\begin{array}{cccc}

			(0, 0) & (0, 1) & (1, 0) & (1, 1)
			\\
			(1 - \tilde w_1^*) (1 - \tilde w_2^*) & (1 - \tilde w_1^*) \tilde w_2^* & \tilde w_1^* (1 - \tilde w_2^*) & \tilde w_1^* \tilde w_2^* 
		\end{array}
	\right) .
	\]
	
	Accordingly, the design $\xi^*$ is also optimal for minimization of the asymptotic variance for estimating the $\alpha$-quantile $t_\alpha$ of the failure time for soft failure due to degradation, when $0 < t_\alpha < \infty$. 
	For instance, under the normal use conditions $x_{u1}=-0.40$ and $x_{u2}=-0.20$ the optimal marginal weights are $\tilde w_1^* = 0.222$ and $\tilde w_2^* = 0.143$, and the optimal design $\xi^* = \xi_1^* \otimes \xi_2^*$ is given by
\begin{equation}
\begin{split}
 {{\xi}^*} = \left(\begin{array}{cccc}(0,0)&(0,1)&(1,0)&(1,1)\\ 0.67&0.11&0.19&0.03
\end{array}\right)
 \end{split}
\end{equation}\\
where the indices of the optimal support points in ${{\xi}^*}$ correspond to the design variables $x_1$ and $x_2$, respectively. Sensitivity analysis proved that ${\xi}^*$ is robust against misspecification of the parameter vector $\BS\beta$. To exhibit the dependence on the normal use condition, the optimal weights $w_1^*,w_2^*,w_3^*,w_4^*$ which correpond to the four vertices $(0, 0)$, $(0, 1)$, $(1, 0)$, and $(1, 1)$ of $\xi^*$, respectively, are plotted in Figure~\ref{fig-weight-x-u1} as a function of $x_{u1}$ where all parameters are held fixed to their nominal values in Table~\ref{table:table34df3}. 
It should be noted that similar results are obtained with regards to $x_{u2}$, and omitted for brevity.
As depicted in Figure~\ref{fig-weight-x-u1} the optimal weights $w^*_3$ and $w^*_4$ which correspond to the maximum testing setting of the first design variable degenerate to zero when the normal use condition approaches the lower bound of $\mathcal{X}_1$, i.e. $x_{u1}\to 0$. 
\begin{figure}[!tbp]
  \centering
  \begin{minipage}[b]{0.44\textwidth}
    \includegraphics[width=\textwidth]{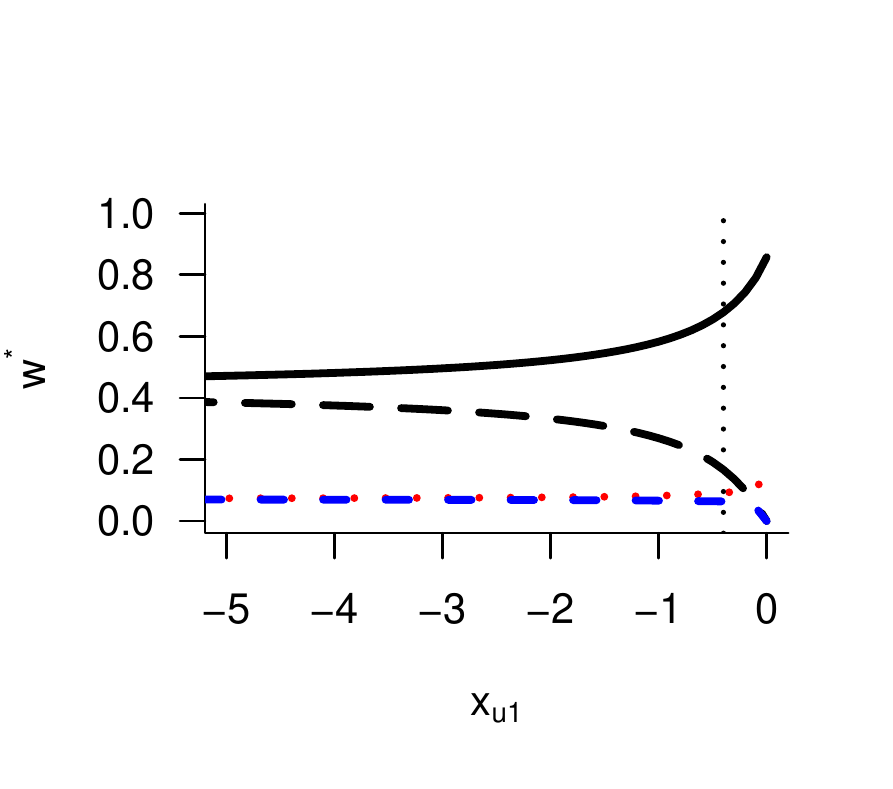}
    \caption{Optimal weights in dependence on $x_{u1}$ for for Example~\ref{example-with-intercept}, solid line: $w^*_1$, dotted line: $w^*_2$, long-dashed line: $w^*_3$, dashed line: $w^*_4$} 
\label{fig-weight-x-u1}
  \end{minipage}
  \hfill
  \begin{minipage}[b]{0.43\textwidth}
 \includegraphics[width=\textwidth]{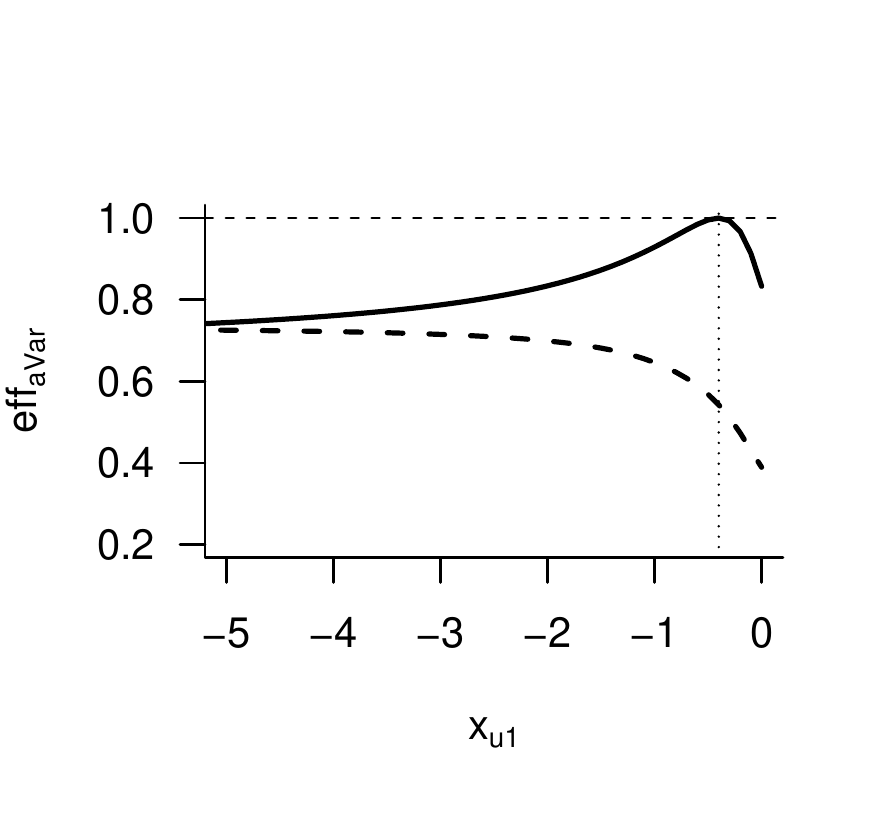}
    \caption{Efficiency of $\xi^*$ (solid line) and $\bar\xi$ (dashed line) in dependence on $x_{u1}$ for Example~\ref{example-with-intercept}}
\label{eff-x1-intercept}
  \end{minipage}
\end{figure}
For the setting of the present model with two stress variable, we examine the efficiency of the design $\xi^* $ which is locally optimal for estimation of the median failure time under the nominal values of Table~\ref{table:table34df3} when the nominal values of $ x_{u1}$ are changed.
	In Figure~\ref{eff-x1-intercept} 
we plot the efficiency of the locally optimal design $\xi^*$ at the nominal values (solid line) together with the efficiency of the design $\bar\xi$ (dashed line) which assigns equal weights $1/4$ to the four vertices $(0, 0)$, $(0, 1)$, $(1, 0)$, and $(1, 1)$ 
where the design $\bar\xi$ is a standard experimental designs for comparison.
In Figure~\ref{eff-x1-intercept} the efficiency is illustrated in dependence on the value of  $x_{u1}$ while all remaining parameters and constants are held fixed to their nominal values in Table~\ref{table:table34df3}. 
	The value for $x_{u1}$ is indicated by vertical dotted lines in the corresponding figures.
In total, the design $\xi^*$ seems to perform quite well and is preferred over $\bar\xi$ throughout. 

\end{example}

\begin{example}
\label{model formulation in randomeffect}

 In this example we consider the model in subsection \ref{multivariaterandom234589} to attain a locally $c$-optimal design for an $s$-out-of-$r$ system with uncorrelated components in the random effects for intercept and slope. The resulting optimal design is attained with regards to the time plan $\boldsymbol{t}=(0, 0.5, 1)^T$ which is unified for all testing units.
We assume here, again, that each of the marginal degradation paths are influenced by two standardized accelerating stress variables $x_1$ and $x_2$ which are defined over the design region $\mathcal X=[0,1]^2$. For some testing unit $i$, the stress variables are set to $x_{i 1}$ and $x_{i 2}$ and the response $y_{i jl}$ of the response component $l$ at time $t_j$ is given by
\begin{equation} 
\begin{split}
\label{example2_2}
 y_{i jl} = &\beta_{il 1} + \beta_{l2} x_{i 1} + \beta_{l3} x_{i 2}+ \beta_{il 4} t_j + \beta_{l5} x_{i 2} t_j + \varepsilon_{i jl},\\ 
=& \mathbf f_l(\mathbf x_i, t_j)\BS \beta_{il}+ \varepsilon_{i jl}\\
\end{split}
\end{equation}
where $\mathbf f_l(\mathbf x, t)=\mathbf f(\mathbf x, t)=(1,x_{1},x_{2},t,x_{2}t)^T$, $\BS \beta_{il}=(\beta_{il 1},\beta_{l 2},\beta_{l 3},\beta_{il 4},\beta_{l 5})^T$, and $\mathbf g_l=\mathbf g=(1,t)^T$.
On the basis of the marginal distribution functions $F_{T_l}(t)$ which defined in \eqref{multivariate-eq-degr-path-use-cond}, the model is extended in terms of the general model \eqref{F-t-marginal} under normal use conditions. 
The aggregate degradation path under normal use condition for response component $l$ is given by
\begin{equation}
\label{marginalfixeddddddd}
 \mu_l(t) =  \mathbf{f}_l(\mathbf{x}_{u}, t)^{T} \BS{\beta}_{l}= \delta_{l1}+ \delta_{l2}t 
\end{equation}
where $\delta_{l1}=\beta_{l1}+\beta_{l2}x_{u1}+\beta_{l3}x_{u2}$ and $\delta_{l2}=\beta_{l4}+\beta_{l5}x_{u2}$ are the intercept and the slope of the aggregate degradation path $\mu_l(t)$ under normal use condition, respectively. 
In the current example we consider the particular case when $r=3$ and $s=2$.
Subsequently, based on equation (\ref{eqn:12}), the joint failure time  distribution for the particular case of a $2$-out-of-$3$ 
can be expressed as 
\begin{equation}
\begin{split}
\label{jointCDF2outof3333}
F_{T}(t)=&F_{T_1}(t)\,F_{T_2}(t)+F_{T_1}(t)\,F_{T_3}(t)+F_{T_2}(t)\,F_{T_3}(t)
-2F_{T_1}(t)\,F_{T_2}(t)\,F_{T_3}(t).\\
\end{split}
\end{equation}
For further calculation we assume that $\rho_l=0, \sigma_{l1}=\sigma_1, \sigma_{l2}=\sigma_2, l=1,2,3$, and, hence, the variance covariance matrix $\BS\Sigma_\gamma$ is a block diagonal matrix with diagonal blocks $\BS\Sigma_{\gamma_0}=\left(\begin{array}{cc}
	\sigma_1^2 &0
	\\
	0& \sigma_2^2
\end{array}\right)$

The distribution function $F_T(t)$ is plotted in Figure~\ref{joint-failure-distrib-bivariatelinear-randomeffect} under the nominal values given in Table~\ref{table:table34df3-random-effect} and the median failure time $t_{0.5} = 2.43$ is indicated by a dashed vertical line. The normal use conditions correspond to $x_{u1}=-0.50$ and $x_{u2}=-0.40$, and the failure thresholds $y_{10}=7.5$, $y_{20}=5.2$ and $y_{30}=4.25$.
\begin{figure}[!tbp]
  \centering
  \begin{minipage}[b]{0.42\textwidth}
 \includegraphics[width=\textwidth]{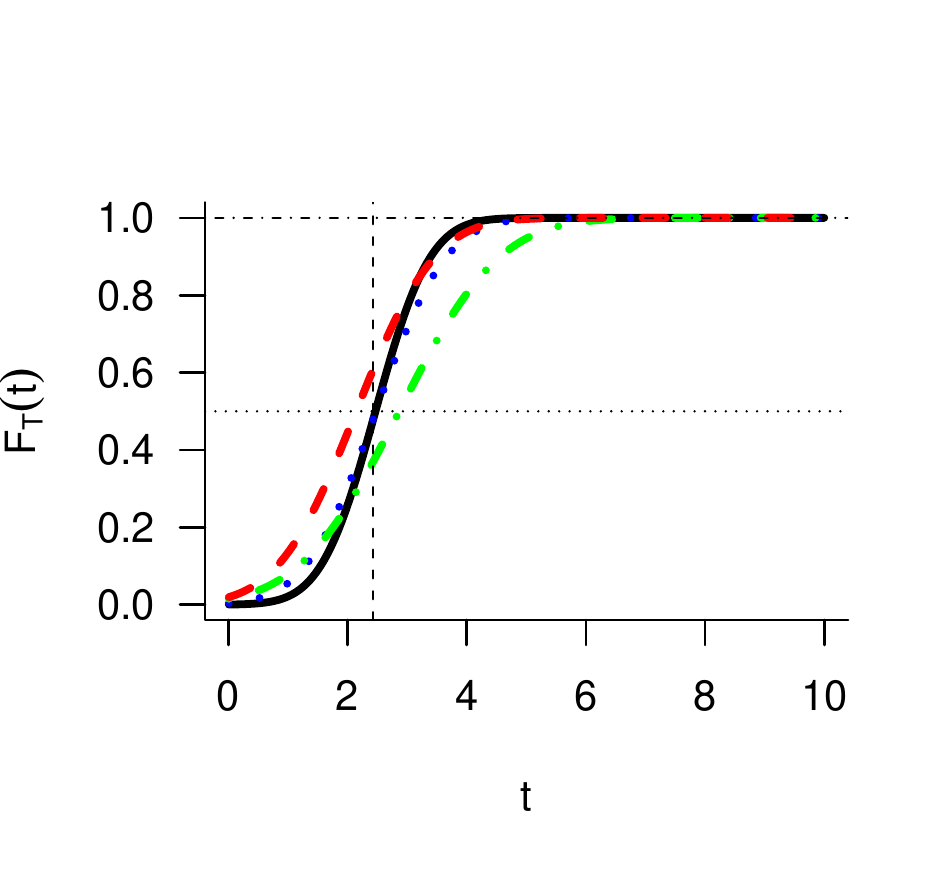}
	\caption{Distribution function $F_T(t)$ (solid line) for the model in Example~\ref{model formulation in randomeffect},  dotted line:$F_{T_1}(t)$,  dotted-dashed line:$F_{T_2}(t)$, dashed line:$F_{T_3}(t)$ } 
	\label{joint-failure-distrib-bivariatelinear-randomeffect}
  \end{minipage}
  \hfill
 \begin{minipage}[b]{0.44\textwidth}
     \includegraphics[width=\textwidth]{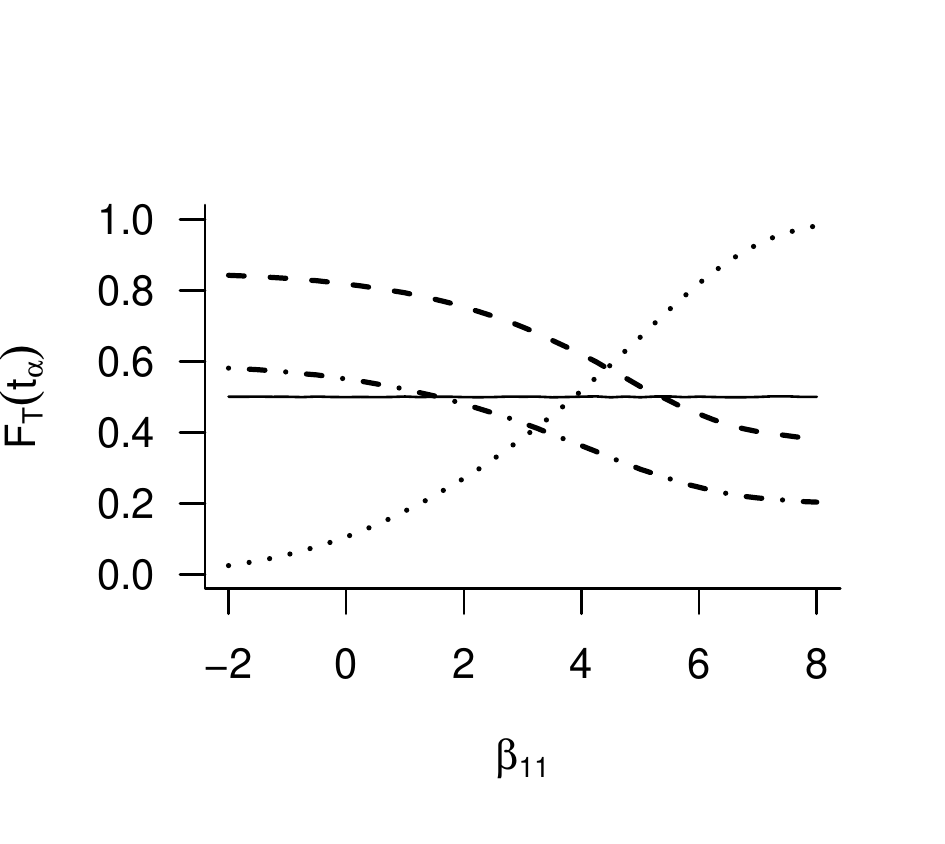}
   \caption{Dependence of $F_{T_1}(t_\alpha)$ (dotted line), $F_{T_2}(t_\alpha)$ (dotted-dashed line) and $F_{T_3}(t_\alpha)$ (dashed line) on $\beta_{11}$ for the model in Example~\ref{model formulation in randomeffect}.}
\label{contribution_b11}
  \end{minipage}
\end{figure}
In view of (\ref{gradieggfgfgf}) and (\ref{jointCDF2outof3333}) the gradient vector of the parameter vector $\BS\beta$ can be expressed as $\mathbf{c}_{\BS\beta}=({c}_{{1}}\mathbf f(\mathbf x_u,t_{{\alpha}})^T,{c}_{{2}}\mathbf f(\mathbf x_u,t_{{\alpha}})^T,{c}_{{3}}\mathbf f(\mathbf x_u,t_{{\alpha}})^T)^T$ where the constants $c_l$, $l=1,...,r$ for a general $s$-out-of-$r$ system are given by
\begin{equation}
\sigma_l(t_\alpha)^{-1} \phi((\delta_{l1} + \delta_{l2} t_\alpha - y_{l0}) / \sigma_l(t)) \sum_{m=0}^{s-1} (-1)^{m}  \binom{m + r - s}{m} \sum_{D: |D|=m+r-s, \,l\not\in D} \prod_{d\in D} F_{T_d}(t_\alpha)
\end{equation}

 and, hence, constants ${c}_{{1}}$, ${c}_{{2}}$ and ${c}_{{3}}$ for the current $2$-out-of-$3$ system are expressed as
\begin{eqnarray*}
\label{multivariate-gradient_cb1-randomeffect}
 {c}_{{1}}&=&\sigma_1(t_\alpha)^{-1}\phi\left(\frac  {\delta_{11}+\delta_{12}t_{\alpha} - y_{10}   }{\sqrt{\sigma_{1}^{2}  + \sigma_{2}^{2} t_\alpha^2}}\right)\left(F_{T_2}(t_\alpha)+F_{T_3}(t_\alpha)-2F_{T_2}(t_\alpha)F_{T_3}(t_\alpha)\right),
\\
 {c}_{{2}}&=&\sigma_2(t_\alpha)^{-1}\phi\left(\frac  {\delta_{21}+\delta_{22} t_{\alpha}- y_{20}   }{\sqrt{\sigma_{1}^{2}  + \sigma_{2}^{2} t_\alpha^2}}\right)\left(F_{T_1}(t_\alpha)+F_{T_3}(t_\alpha)-2F_{T_1}(t_\alpha)F_{T_3}(t_\alpha)\right),\\
 {c}_{{3}}&=&\sigma_3(t_\alpha)^{-1}\phi\left(\frac  {\delta_{31}+\delta_{32}t_{\alpha} - y_{30}   }{\sqrt{\sigma_{1}^{2}  + \sigma_{2}^{2} t_\alpha^2}}\right)\left(F_{T_1}(t_\alpha)+F_{T_2}(t_\alpha)-2F_{T_1}(t_\alpha)F_{T_2}(t_\alpha)\right).\\
\end{eqnarray*}


\begin{table}
	\begin{center}
		\caption{Nominal values of the multivariate linear model with random effect.}
		\label{table:table34df3-random-effect}
		\vspace{5mm}
		\begin{tabular}{c|c|c|c|c||c}

			 $\beta_{11}$ & $\beta_{12}$& $\beta_{13}$ & $\beta_{14}$& $\beta_{15}$ & $\sigma^2_\varepsilon$   \\ \hline
$3.80$ & $0.52$& $0.72$ & $2.00$& $0.67$ &$0.15$  \\ \hline 

\hline \hline

 $\beta_{21}$ & $\beta_{22}$& $\beta_{23}$ & $\beta_{24}$& $\beta_{25}$ & $\sigma^2_1$  
			\\ \hline 
$2.20$ & $0.44$& $0.64$ &$1.50$ & $0.63$ 
			 & $0.40$ 
 \\ \hline 

\hline \hline

 $\beta_{31}$ & $\beta_{32}$& $\beta_{33}$ & $\beta_{34}$ & $\beta_{35}$ & $\sigma^2_2$ 
			\\
			\hline
 $1.33$ & $0.30$& $0.92$ &$1.91$ & $0.80$ 		
&$0.32$\\
		\end{tabular} 
	\end{center}
\end{table}
As mentioned earlier in this example the marginal response components are independent and have the same model equation. Hence, in accordance with Example~\ref{example-with-intercept}, the optimization is reduced to finding an optimal design of the first response component under the normal use conditions. It should be further noted that the resulting locally $c$-optimal design will be optimal for any $s$-out-of-$r$ system under the assumption of  independent response components with the same model equation. In other words, under the assumptions $\mathbf f_l=\mathbf f,\,\, \mathbf g_l=\mathbf g,\,\, \Sigma_{\gamma_l}=\Sigma_{\gamma_0}$ the $c$-optimal design for extrapolation at $(\mathbf x_u,t_\alpha)$ in the LMEM with fixed time plan $\mathbf t$ is optimal for estimating $t_\alpha$.
 In contrast to Example~\ref{example-with-intercept}, it should be mentioned that the optimal design for the current experimental settings depends on the given time plan $\BS t$ as well as the nominal values of $\BS\beta$, through the value of $t_\alpha$,
 due to the particular form of the gradient $c_{\BS\beta}$ as well as the degradation path in equation\eqref{example2_2}. In particular, the optimal design may also vary with $\alpha$ in contrast to the situation in Example~\ref{example-with-intercept}.
In order to derive a locally $c$-optimal design $\xi^*$ that minimizes the asymptotic variance of  $\widehat t_{0.5}$, the multiplicative algorithm (see e.g. \citep{silvey1978algorithm} ) with a grid of marginal 0.05 increments over the standardized design region $\mathcal X=[0,1]^2$ is used. 
The resulting optimal design is given by
\begin{equation}
\begin{split}
\label{optaim2outof3sdglkjhfdg37465}
 {\xi}^* = \left(\begin{array}{cccc}(0,0)&(0,1)&(1,0)&(1,1)\\ 0.60&0.03&0.13&0.24
\end{array}\right)
 \end{split}
\end{equation}
where the general equivalence theorem is applied to prove the optimality of the 
numerically obtained design on the extremal points of the design region. \citep{atkinson2007optimum} state that if $\xi^*$ is an optimal design, the general equivalence theorem insures, under the assumption that the objective function $\Phi$ is a convex function (on the set of all positive definite matrices), that the following three statements are equivalent.
\begin{enumerate}
    \item  The design $\xi^*$ minimizes $\Phi(\mathbf{M}_{\BS\beta}(\xi))$,

    \item  The design $\xi^*$
maximizes the minimum over $\mathcal X$ of $\BS\Psi(\xi,\xi_{ \mathbf x})$,

    \item  The minimum over $\mathcal X$ of $\BS\Psi(\xi^*,\xi_{ \mathbf x})$ is equal to zero,
\end{enumerate}
where $\BS\Psi(\xi,\xi_{ \mathbf x})$ is the directional derivative at the approximate design $\xi$ in the direction of the design $\xi_{ \mathbf x}$  which puts unit mass at the setting ${\mathbf x}$.
 The locally optimal designs for estimating the median failure time are influenced by the parameter vector $\BS\beta$ as well as the normal use conditions $\mathbf x_u$. For brevity, we consider $\beta_{11}$ and $x_{u1}$ for further analysis procedures.
Sensitivity analysis procedures are conducted to demonstrate how the optimal designs change with the parameters and how well they perform under variations of the nominal values. The optimal weights $w_1^*, ..., w_4^*$ which correpond to the four vertices $(0, 0)$, $(0, 1)$, $(1, 0)$, and $(1, 1)$ in \ref{optaim2outof3sdglkjhfdg37465}, respectively, are depicted in Figure~\ref{optimalweightvstimequantile} as a function of $\beta_{11}$ where the variations of $t_{0.5}$ have been generated by letting $\beta_{11}$ vary over the range $-2$ to $5$ and fixing all remaining parameters to their nominal values in Table~\ref{table:table34df3-random-effect}.
The analysis indicated that the optimal weights in \eqref{optaim2outof3sdglkjhfdg37465} slightly change under variations of $\beta_{11}$. 
On the other hand the optimal weights $w_1^*, ..., w_4^*$ are plotted in 
Figure~\ref{fig-weight-x-u-1} as a function of $x_{u1}$, while all remaining parameters are held fixed to their nominal values in Table~\ref{table:table34df3-random-effect}. The results exhibit that the optimal weights are more sensitive to variations of $x_{u1}$ when compared to the misspecifications of $\beta_{11}$. Further, Figure~\ref{contribution_b11} illustrates the dominance of the marginal failure components where the marginal distribution functions $F_{T_{1}}(t_\alpha)$, $F_{T_{2}}(t_\alpha)$, and $F_{T_{3}}(t_\alpha)$ are shown in dependence on $\beta_{11}$. Figure~\ref{contribution_b11} depicts that the first component dominates for 
large values of its intercept $\beta_{11}$ while the second and third components dominate for small values of $\beta_{11}$. 
\begin{figure}[!tbp]
  \centering
  \begin{minipage}[b]{0.42\textwidth}
   \includegraphics[width=\textwidth]{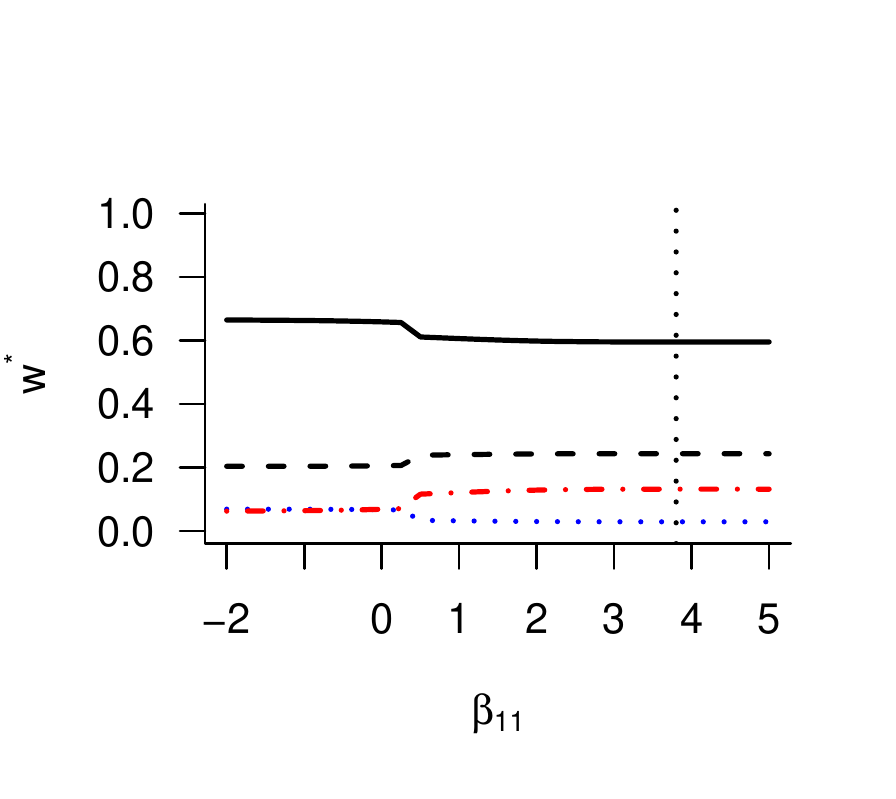}
    \caption{Optimal weights in dependence on $\beta_{11}$ for the model in Example~\ref{model formulation in randomeffect}, solid line: $w^*_1$, dotted line: $w^*_2$, dotted-dashed line: $w^*_3$, dashed line: $w^*_4$.}
\label{optimalweightvstimequantile}
  \end{minipage}
  \hfill
 \begin{minipage}[b]{0.44\textwidth}
   \includegraphics[width=\textwidth]{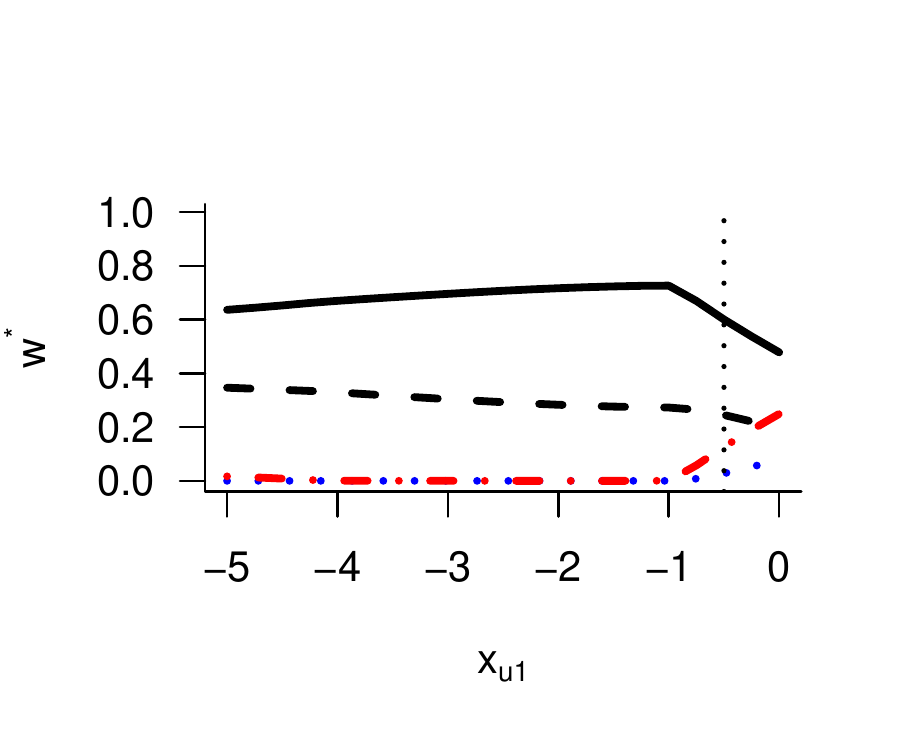}
    \caption{Optimal weights in dependence on $x_{u1}$ for the model in Example~\ref{model formulation in randomeffect}, solid line: $w^*_1$, dotted line: $w^*_2$, dotted-dashed line: $w^*_3$, dashed line: $w^*_4$.}
\label{fig-weight-x-u-1}
  \end{minipage}
\end{figure}
\begin{figure}[!tbp]
  \centering
  \begin{minipage}[b]{0.43\textwidth}
   \includegraphics[width=\textwidth]{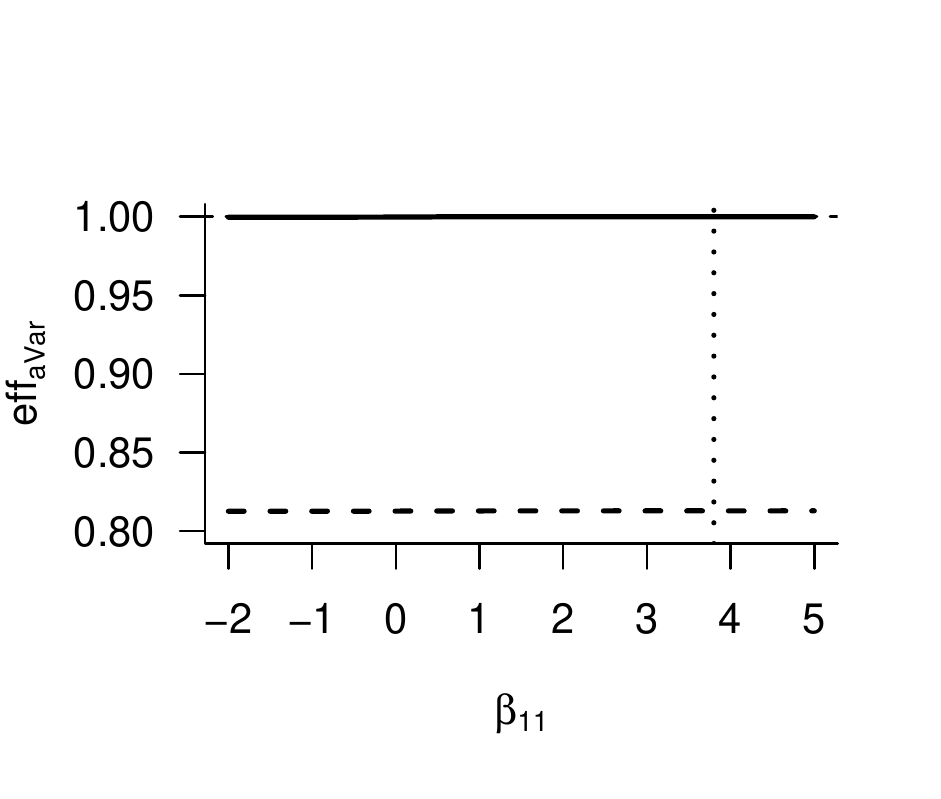}
    \caption{Efficiency of $\xi^*$ (solid line) and $\bar\xi$ (dashed line) in dependence on $t_{0.5}$ for the model in Example~\ref{model formulation in randomeffect}.}
\label{eff-b11}
  \end{minipage}
  \hfill
 \begin{minipage}[b]{0.43\textwidth}
     \includegraphics[width=\textwidth]{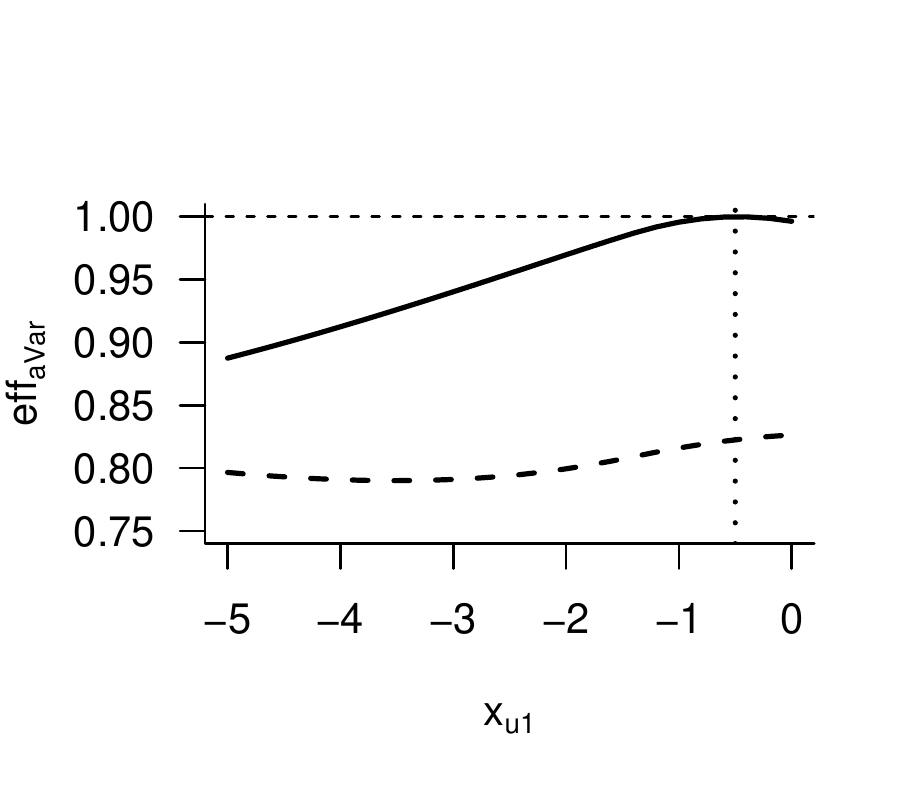}
   \caption{Efficiency of $\xi^*$ (solid line) and $\bar\xi$ (dashed line) in dependence on $x_{u1}$ for the model in Example~\ref{model formulation in randomeffect}.}
\label{eff-x-u-1}
  \end{minipage}
\end{figure}
For the present settings of the mixed effects-model with two stress variable we examine further, based on equation \eqref{eff-forboth}, the efficiency of the design $\xi^* $ which is locally optimal for estimation of the median failure time under the nominal values of Table~\ref{table:table34df3-random-effect} when the nominal values are misspecified.
	In Figure~\ref{eff-b11} and 
Figure~\ref{eff-x-u-1} the efficiency of $\xi^*$ along with the efficiency of $\bar\xi$ are displayed in dependence on the true value of $\beta_{11}$ and the normal use condition $x_{u1}$, respectively. The results indicate that $\xi^*$ performs generally well under misspecification $\beta_{11}$ and $x_{u1}$ with more robustness with regards to variations of $\beta_{11}$.
In total, the optimal design $\xi^*$ is quite preferable over the standard design $\bar\xi$ throughout.
\end{example}

\section{Conclusion}
\label{conclusion}
Designing highly reliable systems needs a sufficient assessment of the reliability related characteristics. A common approach to handle this issue is to conduct accelerated degradation testing which provides an estimation of lifetime and reliability of the system under study in a relatively short testing time. 
To account for variability between units in accelerated degradation tests, we assume int this work that the marginal degradation functions can be described by a mixed-effects linear model.
This also leads to a non-degenerate distribution of the failure time, due to soft failure by exceedance of the expected (conditionally per unit) degradation path over a threshold, under normal use conditions.
Therefore we are aiming to estimate certain quantiles of the joint failure time distribution as a property of the reliability of the product.
In this regard we considered the availability of non-degenerate solutions for the quantiles.
The purpose of optimal experimental design is then to find the best settings for the stress variables to obtain most accurate estimates for these quantities.

For the existing degradation models in this work it is further assumed that stress remains constant within each testing unit during the whole period of experimental measurements but may vary between units.
Hence, in the corresponding experiment a cross-sectional design between units has to be specified for the stress variable while for repeated measures.

In the present paper we presented optimal designs for accelerated degradation testing under bivariate LMEMs with full as well as partial interactions between the time and stress variables.

For all models the efficiency of the corresponding optimal design is considered to assess its performance when nominal values are varied at the design stage.

The construction of designs which are robust against misspecification of the nominal values, such as maximin efficient or weighted (``Bayesian'') optimal designs are object of further research.

\section*{Acknowledgement}
This work has been supported by the German Academic Exchange Service (DAAD) under grant no.~2017-18/ID-57299294.

\end{document}